# Determining Gas Composition for Growth of BNNTs Using Thermodynamic Approach*

Alexander Khrabry[1,2], Igor D. Kaganovich[1], Shurik Yatom[1], Vlad Vekselman[1], Jelena Radić-Perić[3], John Rodman[1,4], Yevgeny Raitses[1]

## Abstract

A high-yield production of high-quality boron-nitride nanotubes (BNNTs) was reported recently in several publications. A boron-rich material is evaporated by a laser or plasma in a nitrogen-rich atmosphere to supply precursor gaseous species for nucleation and growth of BNNTs. Either hydrogen was added or pressure was increased in the system to achieve high yield and high purity of the synthesized nanotubes. According to the widely-accepted "root grow" mechanism, upon the gas cooling, boron droplets form first, then they adsorb nitrogen from surrounding gas species, and BNNTs grow on their surfaces. However, what are these precursor species that provide nitrogen for the growth is still an open question. To answer this question, we performed thermodynamic calculations of B-N mixture composition considering broad set of gas species. In enhancement of previous studies, the condensation of boron is now taken into account and is shown to have drastic effect on the gas chemical composition. $B_2N$ molecules were identified to be a major source of nitrogen for growth of BNNTs. Presence of $B_2N$ molecules in a B-N gas mixture was verified by our spectroscopic measurements during a laser ablation of boron-rich targets in nitrogen. It was shown that the increase of pressure has a quantitative effect on the mixture composition yielding increase of the precursor density. The hydrogen addition might open an additional channel of nitrogen supply to support growth of BNNTs. Nitrogen atoms react with abundant $H_2$ molecules to form $NH_2$ and then $NH_3$ precursor species, instead of just recombining back to inert $N_2$ molecules, as in the no-hydrogen case. In addition, thermodynamics was applied in conjunction with agglomeration theory to predict the size of boron droplets upon growth of BNNTs. Analytical relations for identification of crucial species densities were derived.

[1] Princeton Plasma Physics Laboratory, Princeton University, NJ, USA

[2] Current affiliation: Lawrence Livermore National Laboratory (LLNL). LLNL is operated by Lawrence Livermore National Security, LLC, for the U.S. Department of Energy, National Nuclear Security Administration under Contract DE-AC52-07NA27344

[3] University of Belgrade, Serbia

[4] University of Syracuse, NY, USA

# I. Introduction

Boron-nitride nanotubes (BNNTs) attract significant interest because of their unique properties, such as high Young's modulus[1,2], excellent thermal conductivity[3], superhydrophobicity[4], relatively lower rigidity in the transverse direction[5] and chemical stability up to 900°C in air. Single-walled BNNTs are wide band gap semiconductors with a very small dependence on the chirality[6]. Carbon doping of BNNTs offers a possibility of tailored electrical properties of a nanotube[7,8]. BNNTs can potentially be applied for oncology therapies[9,10] and for water desalination[11,12]. There is an increasing amount of works devoted to synthesis of BNNTs. A thorough review on BNNTs production methods, properties and applications can be found in Ref. [13].

Successful high-selectivity production of high-quality BNNTs was reported in atmospheric and higher pressure systems, including the arc ablation[14,15] and laser ablation[16,17,18,19] reactors and, more recently, in inductively-coupled plasma (ICP) systems[20,21,22]. The later offers scalable high-yield production of BNNTs at relatively low cost. Noteworthy that these methods enabled production of BNNTs without addition of any metallic catalysts to the synthesis process, unlike other methods[23,24]. In both these methods, solid boron or boron-rich material in a bulk or powder form is evaporated in nitrogen atmosphere by a laser or high-temperature ICP plasma forming B-N atomic gas mixture at high gas temperatures. In colder regions, the gas condenses resulting in formation of nanoparticles. A thorough review on BNNTs synthesis methods can be found in Ref. [25].

A commonly accepted mechanism of the formation of BNNTs from the condensing gas is so-called "root-growth" mechanism, first proposed in Ref. [16] based on results of ex-situ analysis, then supported in Refs. [20] and [21], and later witnessed by ab-initio molecular dynamic simulations and arc synthesis experiments[26]. According to the root-growth concept, when B-N gas mixture cools down, boron gas condenses first and droplets of liquid boron form. Then, boron within the droplets reacts with nitrogen-containing radicals from the ambient gas, and BNNTs grow out from the droplets surfaces. However, what are these gas species that provide nitrogen for the BNNTs growth is still an open question. In Ref. [20], for instance, it is assumed that $N_2$ molecules might be precursors for the growth of BNNTs. However, it is questionable whether $N_2$ molecules can serve as precursors for the formation of periodic (hexagonal) B-N solid structures because this transformation is energetically unfavorable: $N_2$ molecules have high binding energy of 9.8 eV whereas B-N bond energy is about 4.0 eV. Other gaseous species such as atomic nitrogen N and BN, $BN_2$ and $B_2N$ molecules look more favorable precursors for formation of hexagonal BN structures.

In recent molecular dynamics simulations[27], several atomic and molecular gas species were tried as "building blocks" to produce solid BN structures. It was shown that tube-like and cage-like BN structures can be efficiently formed from BN molecules. However, these simulations have been performed with pre-defined gas composition; in particular, pure BN molecules gas and pure borazine gas were separately considered. In practice, the feedstock material is introduced into plasma, evaporates and is subsequently transported to the reaction zone[20,21,28]. In this zone, under definite temperature conditions, boron starts to condense and nanotubes form. In this process, composition of the gas mixture cannot be arbitrary prepared; it is determined by chemical reactions between the gas species. What can be controlled is feedstock material, pressure and temperature within the system and buffer gas composition[29,30,31]. Molecular dynamics simulations are not capable of modeling long time-scale processes of material evaporation, cooling and condensation; different method should be applied to determine actual gas mixture composition at the nanoparticles growth.

Interestingly, for all plasma-based BNNT production methods (laser ablation, arc discharge and ICP plasma) rather low production rates were observed at atmospheric pressure: many of the boron droplets ended up with no nanotubes grown on them. Increase of the pressure up to 10 atm. in Ref. [20] and up to 20 atm. in Ref. [17] resulted in substantially higher production yield of the nanotubes and their purity measured in terms of quantity of the nanotubes to the rest of synthesized product. There were also studies where even higher pressures were tried, up to 68 atm.[32] Another approach to increase the yield of BNNTs was found in Refs. [21] and [22]: it was shown that addition of hydrogen to the working gas can increase the yield and purity of the nanotubes produced even without increase of pressure. It is interesting to investigate the gas mixture compositions in these systems and to identify which gas species are crucial precursors for efficient growth of BNNTs.

Thermodynamic modeling[33,34,35] is well suitable for finding composition of a chemically reacting system close to equilibrium state. There are several studies[36,37] devoted to determining composition of the B-N gas mixture at various temperatures. In Ref. [37], a broad set of gas species was considered including three-atomic molecules $B_3$, $B_2N$ and $BN_2$. Intriguingly, $B_2N$ molecules were shown to be dominant species at lower temperatures corresponding to formation of solid BN structures suggesting them as a main precursor for growth of BNNTs. However, in previous thermodynamic computations[36,37,38], condensation of boron was not taken into account (only gas phase species were considered) resulting in overestimation of boron-containing gas species densities at low temperatures where BNNTs grow.

In this paper, we present the results of thermodynamic calculations for B-N system equilibrium composition at atmospheric and increased pressures with and without addition of hydrogen. A broad set of B, N and H containing gas species from [37] was considered with thermodynamic data from Refs. [39] and [40]. Liquid boron and solid BN were taken into consideration as well. This allowed us to account for condensation of boron and determining the conditions for solid BN formation.

Similar thermodynamic approach was previously used in Refs. [41], [42] and [43] to calculate carbon-helium gas mixture composition with condensation of carbon, and in Refs. [44] and [45] to examine gas mixture composition in a chemical vapor deposition process for production of $MoS_2$ layers and in complex metal hydride systems for storage of hydrogen. In Ref. [42], species density profiles in a carbon arc for synthesis of nanoparticles were computed and compared to in-situ measurements. A good agreement with experimental data on species density profile was obtained in Ref. [42], despite fast flow, sharp density gradients and oscillations of the arc showing that chemistry is much faster than other processes and equilibrium assumption can be applied when determining mixture composition for synthesis applications. Computations results for the B-N system were verified by comparison to optical emission spectroscopy (OES) data from our experiments on laser ablation of boron-rich targets in nitrogen atmosphere and to OES measurements [22] in the ICP plasma reactor.

The paper is organized as follows. In section II, thermodynamic method based on minimization of Gibbs free energy for determining equilibrium composition of chemically reacting mixture containing gas, liquid and solid species is described. Section III is devoted to results of the thermodynamic calculations and their analysis. In subsection III.A, effect of gas condensation on the gas mixture composition is evaluated. In subsections III.B and III.C, effects of pressure variation and hydrogen addition on mixture composition and densities potential precursors for growth of BNNTs are studied. Subsection III.D is devoted to verification of the thermodynamics results via comparison to spectroscopic data for ICP reactor [22]. Subsection III.E describes experimental setup used for laser ablation of boron-rich targets and presents OES measurements results. In subsection III.F, the thermodynamics results are used in conjunction of with a simple agglomeration theory to predict size of boron droplets by the time when BNNTs start growing on them and they are solidified. In subsection III.G, analytic relations for densities of major mixture components are derived. Finally, the conclusions are formulated.

## II.     Method for determining equilibrium chemical composition

For mixtures containing few components, a composition can be conveniently determined using thermodynamic equilibrium constants for decomposition and ionization reactions, together with the mass conservation law and electrical neutrality, see Ref. [46] and references therein. For more complex mixtures, as we have here, number of chemical reactions between different species becomes very large, and a different approach, based on the minimization of the Gibbs free energy, is more suitable.

According to the second law of thermodynamics, when a thermodynamic system reaches equilibrium state at given temperature and pressure its Gibbs free energy is at its minimum. Constant pressure is a good approximation when considering any gas volume moving with the flow because Mach numbers in the reactors are normally very low (much less than unity). Typically, temperature variation characteristic time is larger than time required for chemical composition to change, hence, mixture composition can be considered in chemical equilibrium and thermodynamic approach can be applied. Applicability of the thermodynamic approach to determining chemical mixture composition in BNNT production reactors is discussed in detail in Appendix C. Simple description of the Gibbs free energy minimization method for gas mixtures can be found in Refs. [34] and [35].

The Gibbs free energy of a multi-component system (or mixture) can be expressed as a sum of chemical potentials of its individual component species $\mu_i$ multiplied by their quantities $N_i$ (numbers of atoms or molecules) within the closed system (arbitrary chosen volume moving with the fluid flow) under consideration as[47,48]:

$$G = \sum_i N_i \mu_i. \tag{1}$$

For the gas species, chemical potential can be expressed (in ideal gas approximation) as[35,47,48,49]:

$$\mu_i = RT \ln \frac{p_i}{p_0} + G_i^f(p_0, T) = RT\left(\ln x_i^* + \ln \frac{p}{p_0}\right) + G_i^f(p_0, T). \tag{2}$$

Here, $G_i^f(p_0, T)$ is the molar Gibbs energy of the component $i$ in its pure state, at standard pressure $p_0 = 1\, atm.$ and given temperature $T$, or, in other words, it is the Gibbs energy of formation of species $i$ from its constituting elements in their standard states at temperature $T$ (the Gibbs energy of formation of any species in its standard state is commonly considered zero); $p_i$ is its partial pressure; $p$ is pressure in the system $x_i$ is molar fraction of species $i$ among other gaseous species (excluding solid and liquid species):

$$x_i = \frac{N_i}{\sum_{k \in gas\ species} N_k}. \tag{3}$$

For incompressible solid and liquid species, the energy does not depend on pressure, and the expression for chemical potential reduces to:

$$\mu_i = G_i^f(T). \tag{4}$$

Equilibrium composition of a mixture of known gaseous, liquid and solid components can be determined as a set of $N_i$ which gives the minimum value of the Gibbs energy (1) and satisfies constraints of conservation chemical elements of each sort of atoms within the system:

$$\sum_i a_{i,j} N_i = b_j N^*;\ j = 1, 2, ..., m. \tag{5}$$

Here, $m$ is the number of different sorts of atoms (different chemical elements) within the system, $b_j$ is mole fraction of element $j$ within the system, $a_{i,j}$ is number of atoms of sort $j$ in species $i$, and $N^* = \sum_{i \in all}\left(N_i \sum_j a_{i,j}\right)$ is the total number of various atoms in all species within the system. Because densities and molar fractions of species usually are of interest (not absolute numbers of atoms and molecules), $N^*$ can be chosen arbitrary.

Charge conservation needs to be maintained as well:

$$\sum_i Z_i N_i = 0. \tag{6}$$

Here $Z_i$ is the charge number of species $i$; $Z_i = -1$ for electrons.

Substitution of $\mu_i$ defined by (2) and (4) into (1), using (3), yields following function of species quantities $N_i$ to be minimized:

$$RT \sum_{i \in gas} N_i \left( \ln N_i - \ln \left( \sum_{k \in gas} N_k \right) + \ln \frac{p}{p_0} \right) + \sum_{i \in all} N_i G_i^f(p_0, T) \to \min. \tag{7}$$

Here, $N_i$ are absolute quantities of species within the system. Densities and partial pressures of gaseous species $n_i$ can be derived from their molar fractions, $x_i^*$, defined in (3):

$$p_i = x_i \, p; \quad n_i = x_i \frac{p}{kT}. \tag{8}$$

Simultaneous consideration of solid and liquid species allows accurate accounting for condensation/solidification happening in the considered gas volume at constant pressure and slowly changing temperature.

For B-N and B-N-H mixtures we consider broad set of species including various triatomic molecular gases considered in Ref. [37]. Gibbs energies of formation $G_i^f(p_0, T)$ for gases N, $N_2$, $N_3$, B, $B^+$, $B_2$, H, $H_2$, NH, $NH_2$, $N_2H_2$, $N_2H_4$, $NH_3$, BH, $BH_2$, $BH_3$, and $G_i^f(T)$ for liquid/solid B and solid (crystalline) BN were taken from NIST-JANAF reference[39] where they are tabulated as functions of temperature at atmospheric pressure. For the sake of simplicity, only atomic boron ions were considered; ions of other species have very low densities in these equilibrium calculations with the temperature range of interest (below 6 000 K)[36, 37].

Unfortunately, for other gas species observed in plasma and laser ablation of boron-rich targets in inert gas atmosphere, such as $B_2N$[50,51,52], $B_3$[50,53], and $BN_2$ molecules, thermodynamic data is not given in tables [39]. Nevertheless, for these molecules, free energy of their association from constituting atoms (in some other sources also referred as dissociation energy) was computed *ab-initio* in Ref. [40] using quantum chemistry package Gaussian-86[54]. These data allows easy obtaining Gibbs energy of formation for these molecules which can be self-consistently used use with other Gibbs free energy data on species from Ref. [39]:

$$G_i^f(p_0,T) = \sum_j a_{i,j} G_j^f(p_0,T) - \Delta G_i^f(T). \tag{9}$$

Here, $G_i^f(p_0,T)$ is the Gibbs energy of formation of a molecule $i$, $\Delta G_i^f(T)$ is its free energy of association[40], the summation in the right-hand side is performed over various atoms within the molecule, $G_j^f(p_0,T)$ is Gibbs energy of the atoms in gaseous state[39], and again $a_{i,j}$ is number of atoms of sort $j$ in a molecule $i$. Note that in Ref. [40], for molecules having several isomers, e.g. $B_2N$, individual contributions of various isomers are weighted and the global thermodynamic quantity was presented making implementation of the data very convenient.

In principle, even larger molecules constituting of B and N atoms, such as $B_2N_2$, $BN_3$ and $B_3N$ were observed in some of the experiments[51,52]. However, lack of thermodynamic data for these molecules did not allow taking them into consideration. In this regard $B_2N$ can be viewed as a proxy for other larger molecules, similar to what was observed for carbon where many molecular species can be accounted for[41,55].

Additional complication is that the thermodynamic data for BN molecules[39] is given with a very high level of uncertainty. Free energy of formation is calculated based on dissociation energy for which uncertainty is about 70 %: its value varies between 4 eV and 7 eV for the standard conditions, depending on literature. Such ambiguity in the energy of formation would result in substantial uncertainty (of several orders of magnitude) in computed BN molecules density in the equilibrium B-N mixture. Noteworthy that in Ref. [39] rather old literature was considered when constructing the table for BN molecules, 1964 and before. Newer sources [40, 56, 57, 58, 59, 60] are more consistent on the value of BN dissociation energy at standard conditions, giving values varying from 3.9 eV to 4.5 eV. Possible explanation for such variety of the dissociation energy values in old papers can be deducted from Ref. [59] where it is explained that binding (or dissociation) energy of BN molecule depends on whether it is considered to be in a ground state $3\Pi$ or excited state $1\Sigma$. For the ground state, adequate for our temperatures of interest, binding energy is about 4 eV. For our thermodynamic modeling, we calculated free energy of association $\Delta G_i^f(T)$ for BN molecules using thermodynamic data from Ref. [40], to be consistent with data for other molecules. We also tried thermodynamic data for BN from newer source [58], there were some quantitative changes in the BN densities but the picture did not change qualitatively. The energy was calculated from equilibrium constant $K_f$ for the formation reaction: $\Delta G_i^f(T) = -RT \ln K_f$.

## III. Results and discussion

### III.A. Effect of boron condensation on the species densities

Computed composition of B-N mixture (with no hydrogen), is shown in Figures 1 and 2 as a function of temperature. Figure 1 displays effect of boron condensation on species densities. Two computational runs were performed. In one run, single phase calculation was performed – only gas phase species were considered. Their densities are presented partially as solid lines for T>4137 K (4137 K is the boiling point of liquid boron at atmospheric pressure) and dashed lines, for T<4137 K. Different colors represent different species. In another run, liquid boron and solid BN were added to the system allowing accounting for boron condensation and solid BN formation. The results of this run are plotted with solid lines. In both computational runs, mole fractions of boron and nitrogen elements within the mixture were the same, 45% and 55% respectively; pressure is atmospheric. This boron fraction was chosen arbitrary here; the effect of boron fraction will be discussed in more details in the following section. Though full set of boron- and nitrogen-containing species was modeled, only densities of boron-containing species were affected by the condensation. For the sake of keeping plots in the Figure 1 legible, only these species are shown; other species will be shown in following subsections. $BN_2$ molecules have very low densities compared to other species and are not plotted in Figure 1.

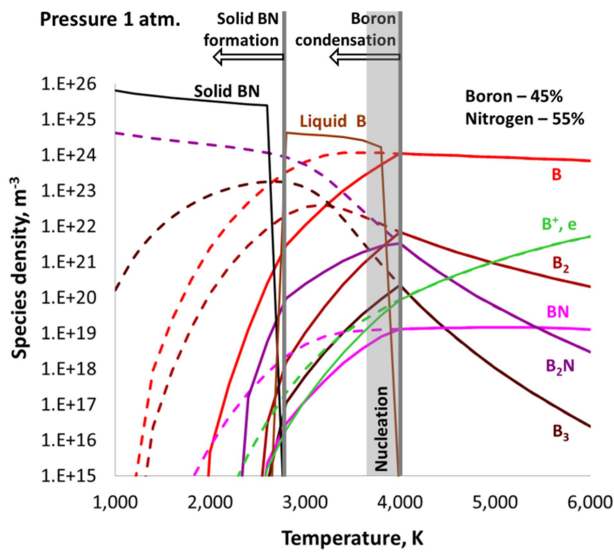

*Figure 1. Effect of boron condensation on gas species densities in the B-N mixture. Solid lines – results obtained with accounting for the condensation; dashed lines – condensation not accounted for. Though full set of species was modeled, only densities of boron-containing species are affected by the condensation and plotted here. Condensation of boron has drastic effect on densities of all boron-containing gas species reducing their density by orders of magnitude. This effect is very important when determining gas species surrounding boron droplets at temperatures of solid BN formation.*

From this figure it is clear that boron condensation has drastic effect on densities of all gas phase boron-containing species. When the gas mixture cools down and temperature reaches 4000 K, thermodynamics predicts that boron starts to condense, see line "Liquid B", which shows density of boron atoms within liquid droplets in a unit gas volume. The condensation at definite temperature, can happen when the system becomes "oversaturated'' in respect to the partial pressure of boron atoms. As a consequence of the formation of liquid boron, the amount of boron in the gas phase is lowered and the equilibrium density ratios among boron-containing species B, BN, $B_2N$, etc. in different gas phase reactions are disturbed. To reestablish this equilibrium, according to the Le Chatelier's principle, the chemical reactions are shifted in direction of the formation of B atoms by additional dissociation of the boron-containing molecules, leading to the further formation of liquid boron until the heterogeneous equilibrium between gas phase boron and liquid boron is completed. This also means that the partial pressure of boron atoms equals to its saturation level value, $p_B^s$ (which is a function of temperature, only); and solid line curve for B at T<4000K denotes the boron saturation vapor pressure curve. That value and thermodynamic equilibrium constants of chemical reactions determine the partial pressure values of other boron containing compounds BN, $B_2N$, $B_2$ etc. This question will be addressed in more detail in subsection III.G, where analytical relations for the species densities as functions of temperature are derived.

When temperature decreases to 3800 K, most of the boron has already condensed. When temperature approaches about 3000 K, thermodynamics predicts that formation of solid BN should begin, see line "Solid BN". Because initially there was slightly more nitrogen than boron in the mixture, see also Figure 2, we can conclude that almost all liquid boron is consumed to form solid BN and some nitrogen is left in form of $N_2$ gas to maintain pressure within the system. Here and on subsequent figures, temperatures corresponding to initiation of boron condensation and formation of solid BN are marked in the figures with vertical grey lines.

Formation of solid BN involves, besides boron liquid, also nitrogen and boron containing gas phase species BN, $B_2N$ etc. Their densities are determined by heterogeneous equilibrium and are considerably lower compared to those computed using one-phase procedure which does not take condensation of boron into account (denoted with dashed lines at T<4000K in Figure 1).

Note that Gibbs free energies[39] for liquid boron and solid BN, which were used in present thermodynamic calculations, were obtained without accounting for surface energy effects for nanoparticles. They are valid for such relatively large volumes of liquid/solid that the surface effects can be neglected. However, at initial stage of condensation, when droplets of liquid boron nucleate and are very small (on a sub-nanometer scale), the surface effect leads to a barrier for the nucleation of droplets[61,62,63]. In the other words, while a gas mixture is cooling down, condensation does not immediately start when temperature reaches saturation point predicted by the thermodynamics. Calculations[64,65,66] for aluminum and our calculations using Nodal General Differential Equation (NGDE) solver[65] and theoretical findings[67] for nucleation of boron droplets have shown that no condensation happens until the actual gas pressure exceeds the saturated vapor pressure predicted by the thermodynamics about five times (i.e., the super saturation degree is about 5 instead of one) with a very weak dependence on a gas cooling rate. This supersaturation degree corresponds to the temperature of about 300 K below the saturation point for boron at atmospheric pressure (due to very strong dependence of vapor pressure on gas temperature). After that, nucleation and growth of boron droplets happens very rapidly, the surface energy effects for the droplets become negligible, the boron vapor condenses onto the droplets, equilibrium between gas and liquid phases is reached, and present thermodynamic predictions of the mixture compositions become valid.

In other words, the actual condensation temperature is about 300 K lower compared to the one predicted by thermodynamic calculations. In a shaded grey region denoted "Nucleation" in Fig. 1, the mixture is well described by the solution with no condensation (dashed lines). At the left boundary of the shaded area mixture composition changes abruptly switching to the solution with condensation (solid lines).

Note that the main focus of this paper is to determine the gas composition at temperatures corresponding to the formation of solid BN structures (e. g. nanotubes), these temperatures (2000-2400K; see Ref. [26]) are substantially lower than the temperature of boron condensation (≈4000 K; see Figure 1). The processes involved in the nucleation of boron droplets are out of the scope of this paper. Therefore, for the sake of simplicity of the subsequent figures, only thermodynamic results of the calculations considering liquid boron (solid lines in current notations) corresponding to gas in equilibrium with liquid are plotted. Assessments presented in Appendix C show that thermodynamic predictions for densities of major mixture components at temperature of solid BN formation are applicable to BNNT production reactors.

### III.B. Effect of pressure increase on the B-N mixture composition

In Figures 2, composition of B-N mixture is shown for pressures 1 atm. and 10 atm. Partial pressures of gas phase species are plotted as functions of temperature. $N_3$ molecules have low densities compared to other species and are not plotted. Temperatures of boron condensation and solid BN formation are shown with grey vertical lines. In order to show effect of boron fraction on the mixture composition, in addition to previously considered boron to nitrogen ratio 45 to 55, the results for a mixture with only 1% of boron are plotted.

Apparently, reduction of the boron fraction affects boron partial pressure at higher temperatures, when no condensation takes place and all species are in the gas phase. As seen from the plots, it results in reduction of temperature at which boron vapor becomes saturated and condensation starts. At temperatures lower than that, fraction of boron in the gas phase and a chemical composition of the gas phase are generally determined only by the saturation vapor pressure and do not depend on the initial fraction of boron within the mixture. This result is also confirmed by analytical derivations presented in section III.G. Amount of boron in the liquid phase is obviously affected by its initial fraction; the initial amount of boron also affects the size of boron droplets formed during the condensation, as will be discussed in section III.F.

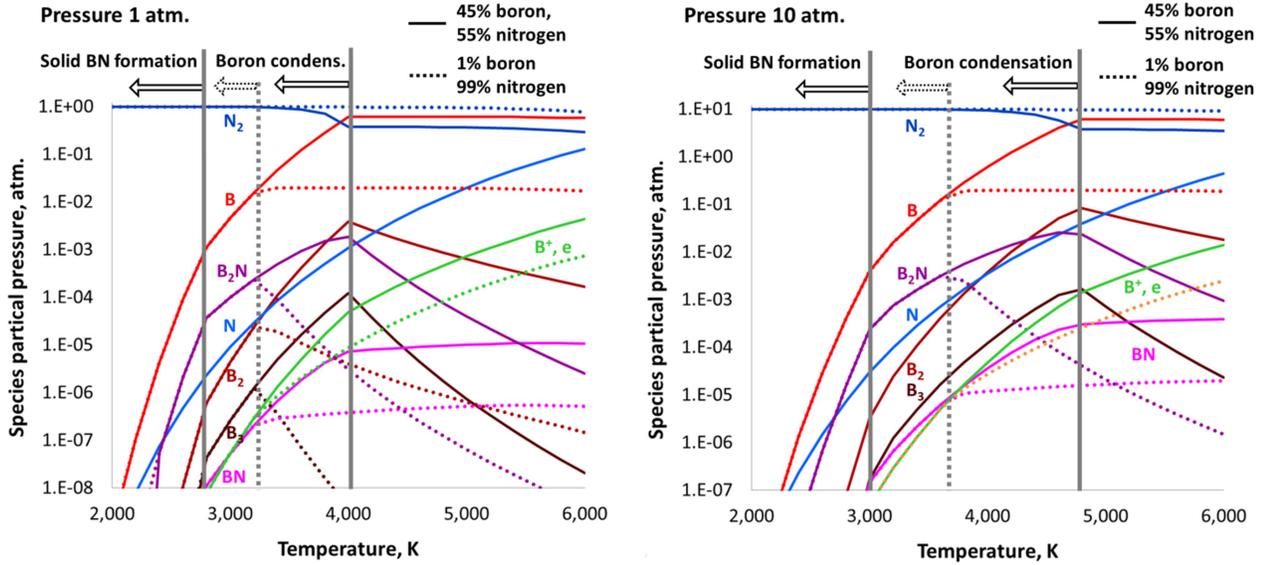

*Figure 2. Composition of B-N mixtures at pressures 1 atm. (left) and 10 atm. (right) having boron to nitrogen ratios 45 to 55 (solid lines) and 1 to 99 (dotted lines). Boron condensation is taken into account; partial pressures of gas phase species are plotted as functions of temperature. Temperatures of boron condensation and solid BN formation are shown with vertical grey lines. Fraction of boron within the mixture affects composition of the gas species only at higher temperatures, before boron condensation starts. When the condensation takes place, gas species composition is determined merely by the boron vapor saturation curve. At the temperature of solid BN formation, $B_2N$ has the highest density among nitrogen containing species, after $N_2$, and can be suggested as major source of nitrogen for formation and growth of BNNTs. With the increase of pressure p, densities of N, BN and $B_2N$ increased as $\sqrt{p}$ (as will be explained in subsection III.G), while the ratios between the species densities remained practically the same.*

When boron starts condensing, molecular nitrogen, $N_2$, becomes the major component in the mixture, with several orders of magnitude higher density than other components. $N_2$ molecules have high dissociation energy of 9.8 eV and can hardly be dissociated at boron droplets (unless they penetrate deep into boron[26]); hence they are unlikely to be effective precursor for formation of B-N nanostructures. Among other nitrogen containing species, $B_2N$ has the highest density, an order of magnitude higher than its closest competitor, atomic N. This is due to its bond energy of about 6eV[40], $B_2N$ molecules can be disassembled at surfaces of boron droplets[26] and can be suggested as major precursor for grows of BNNTs. Atomic nitrogen may contribute to the growth of BN nanostructures as well.

The increase of pressure from 1 atm. to 10 atm. resulted in $B_2N$ partial pressure increase by a factor of 10 at temperature of solid BN appearance, possibly explaining an increase of nanotube production yield observed in Refs. [17] and [20].

### III.C. Effect of hydrogen addition

As proposed in Ref. [21], hydrogen addition could support BNNTs growth via formation of N-H containing gas phase species that could serve as nitrogen supply to boron droplets on which the BNNTs grow (see Fig. 5 in Ref. [21]). In this section, gas phase mixture composition is analyzed to identify these species. Mixtures with 5% and 30% hydrogen addition at atmospheric pressure are considered. Results of the chemical composition computations are presented in Figures 3 and 4. Figure 3 has plots for densities of nitrogen-containing gas species. The rest gas species are shown in Figure 4. Note that $N_2H_2$ and $N_2H_4$ molecules were considered in the simulations but are not plotted because of their negligibly low densities.

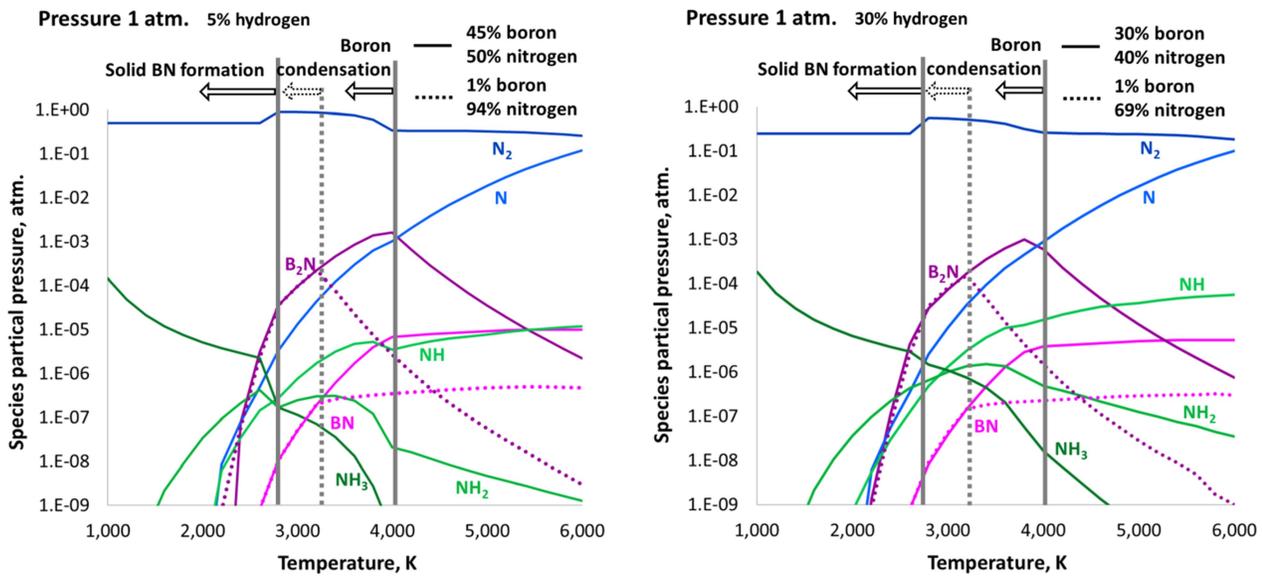

Figure 3. Composition of B-N-H atmospheric pressure mixtures with hydrogen fractions of 5% (left) and 30% (right). Only nitrogen-containing species are plotted here (for other species, see Figure 4). Temperatures of boron condensation and solid BN formation are shown with vertical lines. With high hydrogen fraction, $NH_3$ molecules become prominent near temperature of solid BN appearance, and may serve as a nitrogen source for growth of BNNTs.

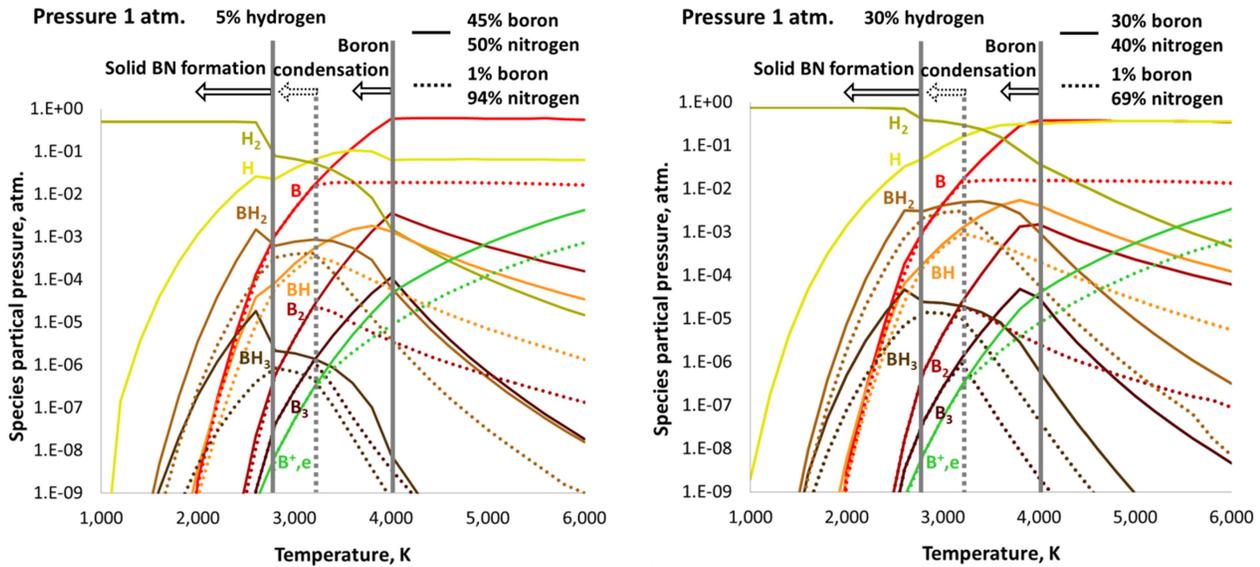

*Figure 4. Composition of B-N-H atmospheric pressure mixtures with hydrogen fractions of 5% (left) and 30% (right). Only species that do not contain nitrogen are plotted (for nitrogen-containing species, see Figure 3). Temperatures of boron condensation and solid BN formation are shown with vertical lines.*

As seen from Figure 3, even with substantial addition of hydrogen (30%), $B_2N$ is still major nitrogen-containing gas phase component (after $N_2$) at the temperature of appearance of solid BN (2800 K). Nevertheless, with further decrease of temperature, densities of all boron-containing species (including $B_2N$) continue to drop rapidly, whereas density of ammonia $NH_3$ molecules grows. Soon, at temperature about 2600 K, $NH_3$ becomes major nitrogen-containing species, after $N_2$. NH and $NH_2$ molecules are also present but have much lower densities than $NH_3$ at these conditions and cannot noticeably affect the growth of BNNTs. As discussed in Appendix C, rates of NH and $NH_2$ destruction and consumption are sufficiently high to reduce their density in accordance with the thermodynamic prediction at temperature of BNNTs growth.

As shown in Refs. [68],[69], molecules of ammonia dissociate at surfaces of boron droplets releasing hydrogen into a gas phase and can serve effectively as feedstock of nitrogen for growth of solid BN structures. $NH_3$ is known to serve effectively as nitrogen source for BNNTs in low temperature (1500 K) furnace reactors[70,71]. It is plausible that $NH_3$ molecules provide nitrogen for further growth of BNNTs at lower temperatures, when amount of $B_2N$ molecules is significantly lowered due to condensation of boron, and boron droplets are not yet solidified, which happens at temperature about 2350 K. In the other words, the addition of hydrogen might open an additional channel of nitrogen supply for growing BNNTs via formation of $NH_3$ molecules. This might explain a higher yield production of longer nanotubes observed in the case of hydrogen addition[22].

Even though it is energetically favorable, as predicted by the thermodynamics, a common concern regarding ammonia formation from $H_2$ and $N_2$ is that the process is known to be very slow at low temperatures due to the high activation energy barrier associated with the need to dissociate molecular nitrogen[72]. In a well-known Haber-Bosch process[72] invented in the beginning of 20$^{th}$ century, metallic (iron) catalyst and high pressure of about 200 atm. are required in order to speed up the $N_2$ dissociation reaction and to produce high yield of $NH_3$; in experiments[73] ammonia is produced during $H_2$ desorption in metal hydride systems. But, in the high-yield BNNT synthesis systems, there are no metallic catalysts; $N_2$ molecules do not dissociate on liquid boron surface[26]. Can $NH_3$ be produced there at sufficient rate? Note that the Haber-Bosh process and experiments[73] were run at constant temperature of 800 K, whereas in nanosynthesys reactors, BNNTs grow at temperature above 2300 K, and the mixture components originate from higher temperature regions with sufficient amount of atomic nitrogen. As we discuss in Appendix C, nitrogen atoms react fast with abundant $H_2$ molecules (much faster than they recombine with scarce each other), forming, in accordance with the thermodynamic prediction, sufficient amount of $NH_2$ and then $NH_3$ molecules in agreement with the thermodynamic prediction. $NH_2$ molecules serve as reaction intermediates; they transform to $NH_3$ at a high rate resulting in considerably higher density of the latest.

### III.D. Comparison to OES data from Ref. [22]

In Ref. [22], high yield production of high quality BNNTs was achieved in ICP plasma reactor at atmospheric pressure with addition of up to 24% hydrogen to the gas mixture. Computational Fluid Dynamics (CFD) Simulations of gas flow with heat transfer were performed. Temperature and velocity profiles were obtained. Evaporation of BN powder and transport of BN vapor were simulated. Complete evaporation of the BN feedstock is suggested. However, chemical reactions within the system were not considered, BN gas was modeled as a non-reacting admixture, and gas mixture composition was not predicted in the modeling. Instead, optical emission spectra were measured at several locations within the reactor.

Simulations results[22] on BN density and temperature allow us to estimate molar fraction of boron within the gas. The data[22] on boron fraction and gas temperature can be used as an input for the thermodynamic modeling to compute the gas mixture compositions at given locations within the reactor. Applicability of equilibrium thermodynamic description to processes in the BNNT synthesis reactor is discussed in Appendix C. These results can be compared to OES data from Ref. [22]. Note that passive spectroscopy was utilized in Ref. [22] which does not provide quantitative data on species densities. Emission of polyatomic molecules is weak, and their spectral lines were not detected at any location within the reactor. Thereby, this experimental data only allows qualitative comparison between densities of atomic/diatomic species predicted by the modeling and their visibility in the emission spectra at various locations within the reactor. The results of this comparison are presented in Table 1. Similar qualitative comparison between the calculations and the experimental results was performed, for instance, in Ref. [45].

First location where the spectrum was measured (0.38 m away from the inlet) had the gas temperature of about 5 000 K. BN gas molar fraction is about 2% meaning about 1% of boron molar fraction. NH molecular emission lines are prominent, N atomic spectral lines and $N_2$ molecular bands were observed as well. Atomic B emission lines were detected as well as emission bands from BH and NH radicals. Spectral lines of atomic H were clearly identified, while the relative weakness of $H_2$ molecular band suggests effective dissociation of $H_2$ into H atoms. A trace of BN molecules was observed but its intensity was much weaker than those of BH or NH radicals. These observations agree with the results of our thermodynamic modeling (see Figure 4). NH molecules have considerable density (molar fraction $10^{-4}$ in case of high fraction of hydrogen within the mixture system). Atomic boron has high density (molar fraction 0.02; almost all boron is in atomic form). Hydrogen is mostly present in atomic form; density of $H_2$ molecules is about 3.5 orders of magnitude lower. Density of BN molecules is two orders of magnitude smaller than those of both BH and NH molecules.

| Location of the spectra measurement; gas temperature | Species detected in the OES spectra[22] | Species predicted by the thermodynamic modeling and their molar fractions |
|---|---|---|
| Distance from the inlet: **0.38 m** | Detected: N, $N_2$, NH, B, BH, H | N ($2\times10^{-2}$), $N_2$ ($2\times10^{-1}$), NH ($3\times10^{-5}$), B ($2\times10^{-2}$), BH ($3\times10^{-5}$), H ($3\times10^{-1}$) |
| Temperature: **5 000 K** | Weak signal: BN, $H_2$ | BN ($3\times10^{-7}$), $H_2$ ($3\times10^{-3}$) |
| Distance from the inlet: **0.48 m** | Detected: BH, H | BH ($3\times10^{-4}$), H ($3\times10^{-1}$) |
| Temperature: **3 700 K** | Not detected: B, NH | B ($2\times10^{-2}$), NH ($10^{-5}$) |
| Distance from the inlet: **0.88 m** | Detected: H | H ($10^{-2}$) |
| | Weak signal: BH | BH ($3\times10^{-6}$) |
| Temperature: **2 400 K** | Not detected: B, NH | B ($3\times10^{-6}$), NH ($3\times10^{-8}$) |

*Table 1. Qualitative comparison between results of the OES measurements[22] and the thermodynamic modeling. Good qualitative agreement is observed between the species densities predicted by the modeling and their visibility by the OES measurements.*

At the next downstream locations where OES measurements were performed (0.48 m and 0.88 m from the inlet) corresponding to temperatures of 3700 K and 2400 K respectively, spectral lines of B atoms and NH molecules were no longer observed, whereas BH molecules and H atoms were still detected.BH molecular band at 0.48 m (3700 K) is the most prominent one. These observations agree well with the thermodynamic results (Figure 4): at temperatures of about 2400 K, density of H radicals is still high; it is of about the same order of magnitude as it was at 5000 K. Whereas, density of boron atom reduces about two orders of magnitude at 2400 K due to condensation of boron. Density of BH molecules at 3700 K is higher than at 5000 K, and at 2400 K is about the same as at 5000 K.

Note that in Ref. [22], enhanced growth of BNNTs in case of hydrogen addition to the gas mixture was attributed to the presence of NH radicals in the mixture. It was assumed that BNNTs start to grow immediately when boron starts to condense, at temperature slightly below 4000 K, where density of NH radicals is high. However, results of thermodynamic modeling suggest that nanotubes and other solid BN structures start growing at temperatures below 2 800 K where density of NH molecules is rather low, which is in agreement with the OES measurements. $NH_3$ molecules have higher density at these conditions, as discussed in a previous section.

### III.E. Laser ablation of boron-rich targets: experimental setup and results

The sketch of the setup and the imaging on the iCCD is presented on Figure 5. The wavelength calibration was conducted with Hg and Xe calibration lamps: resulting diffusion relation was determined to be 0.18-0.16 A/pixel and the instrumental width for slit with of 50 micron is ~1 A. The experiments of laser ablation of BN target were conducted in a chamber in order to control the gas environment. Prior to the experiment the chamber was evacuated and subsequently filled with either He or $N_2$ gas, up to pressure of P=400 Torr. Details on the experimental setup are provided in appendix A.

The identification of atomic species was done according to the NIST database[74] and the molecular species – according to Refs. [75, 76]. Overall we have detected 3 band-heads of $B_2N$ at 488.1, 504.3 and 519.5 nm [76], in experiments of BN target ablation in He and in $N_2$ environments. Figure 6 shows the temporal evolution of the spectrum in range 480-497 nm within the first microsecond after the laser pulse, in the experiment conducted in He-filled chamber. Presented spectra were captured with exposure of 50 ns, in 50 ns intervals, accumulating the emission from 100 shots. In order to show the temporal evolution in a single figure the intensity of each spectrum was normalized in the range (0, 1). The boron and nitrogen ion lines are prevalent in the spectrum, following the laser impact on the BN target. These lines are strongly broadened, owing to the high density and temperature in the spatial-temporal vicinity of the laser impact. After ~300 ns they disappear and the $B_2N$ band emerges. The molecular $B_2N$ emission bands were observed for tens of microseconds, after the laser shot. A more detailed analysis of these results will be described in a separate paper. For the purpose of this work, the most important result is the observation of molecular species in the ablation plume which were also predicted by thermodynamic calculations.

To summarize, spectroscopic measurements verify presence of $B_2N$ molecules in the B-N gas mixture at nearly atmospheric pressure. Interestingly, immediately following the laser pulse the $B_2N$ molecules are not detected, only its precursors (B and N ions and atoms); it takes several fractions of microsecond for $B_2N$ molecules to appear. This behavior supports the point that $B_2N$ molecules do not come directly from the target; they are formed by chemical reactions in the gas phase, as suggested by the thermodynamic modeling.

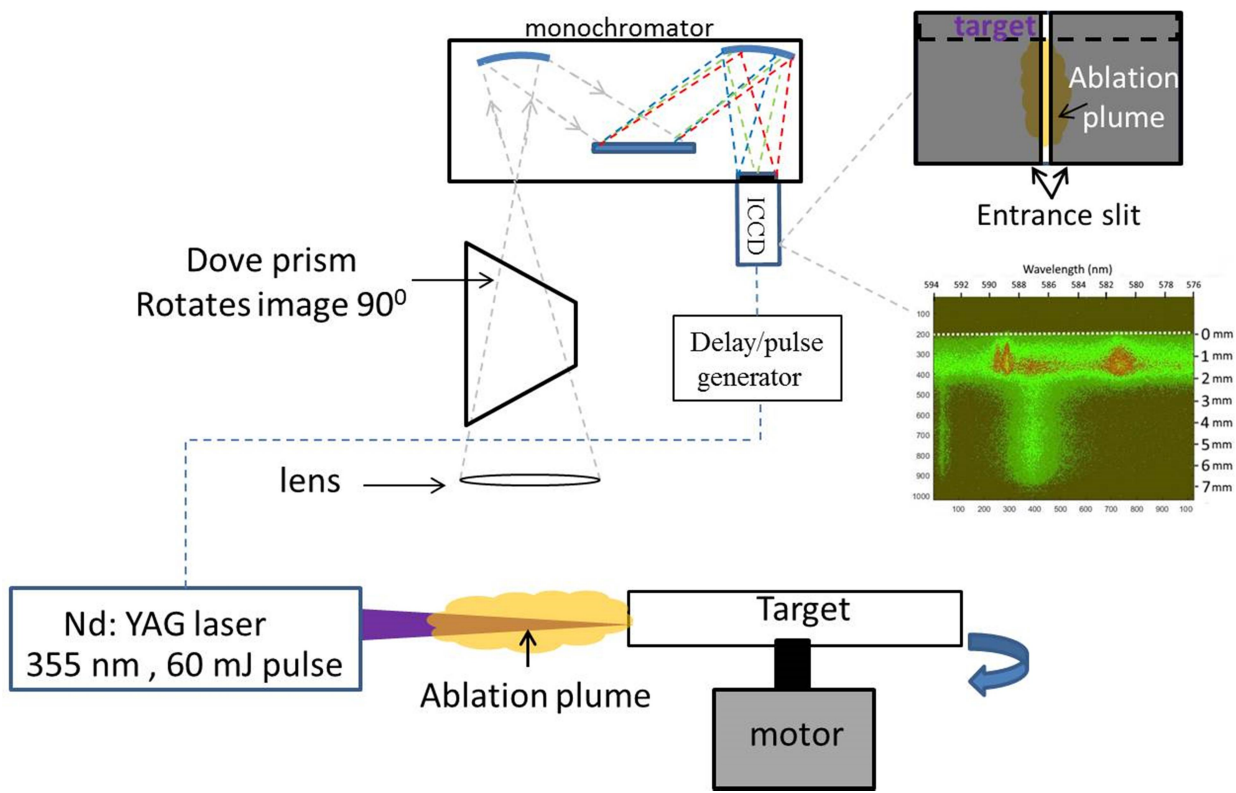

*Figure 5. Experimental setup in laser ablation experiment of BN target.*

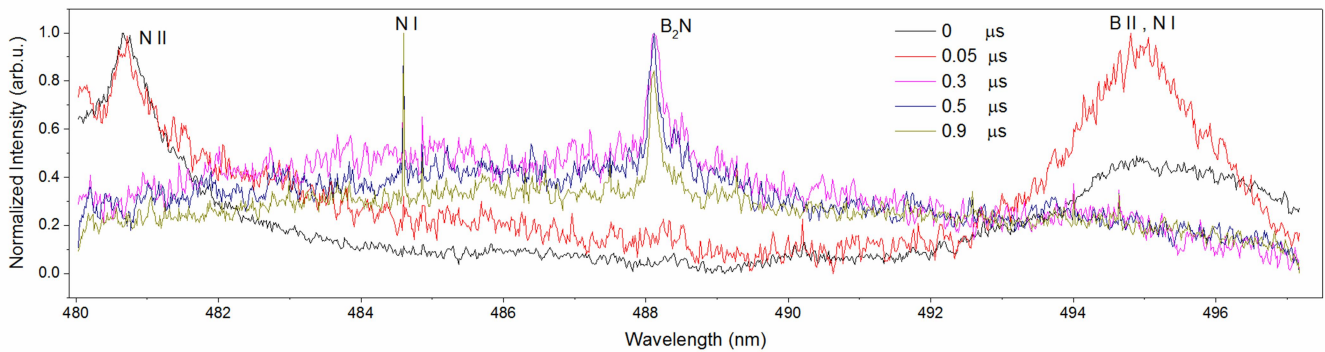

*Figure 6. Temporal evolution of the spatially-integrated, ablation spectrum of solid BN in He (P=400 Torr), in the 480-497 nm range. The intensities were normalized in 0-1 range in order to show the spectra on the single plot.*

### III.F. Estimation of boron droplets size upon formation of BNNTs in an ICP and arc discharge reactors

Experiments[21,77] have shown that diameters of BNNTs are strongly correlated with the sizes of catalyst droplets on which they grow, and formation of nano-sized boron droplets with suppression of their growth is important for the production of small-diameter BNNTs. Similar effect is well known for carbon nanotubes (CNTs)[78]. Unfortunately, not that much research was done for BNNTs as it was done for CNTs. Besides, some sources[15,16] do not report correlation between the droplets and nanotubes diameters. Thorough experimental and theoretical research is required to answer a question if there is a correlation and how to control the properties of nanotubes produced. This research is out of the scope of this paper. In this section, we provide a simple expression for boron droplets size, which can facilitate this research.

Growth of boron droplets can be divided in two stages[61]: 1) the nucleation stage refers to a statistical process of boron atoms aggregation into clusters in which stable boron droplets are formed; 2) the agglomeration stage refers to a process of droplets growth via collisions with each other leading to their coagulation into larger droplets.

In a cooling gas mixture, nucleation of droplets does not happen immediately when conditions for the gas condensation are met as predicted by thermodynamics. This is due to energy barrier to the formation of stable droplets[61]. For boron, the "delay" is about 300 K, with a very weak dependence on the gas cooling rate. When this condition on the temperature is met, nucleation happens abruptly, and many stable droplets are rapidly formed. Once the droplets are formed, their growth is governed by a coagulation (droplets merge when colliding with each other).

If the gas is ionized, particles may become charged, which in turn can affect the coagulation process, as was shown e.g. in Refs. [79], [80]. However, in Refs. [79], [80] and other, low pressure plasma was typically considered, where electron temperature reaches 5 eV (typically maintained by an RF source) while the gas and the particles are at a room temperature. Such electron temperature provides high ionization rate sufficient to compensate the loss of charge particles due to the ion (positive ion-electron) recombination at the droplets surfaces, and maintain the plasma. In atmospheric (and higher) pressure reactors considered in this paper, electron temperature is close to the gas temperature (is about 0.3 eV in the coagulation region). At this temperature, ionization rate is quite low and plasma will be virtually immediately extinguished via interaction with the droplets as soon as they appear, as is shown in details in Appendix B.

Thereby, coagulation will occur in a neutral gas with neutral particles, and a simple agglomeration theory[81] can be employed to describe this process. This theory was applied in Ref. [43] to describe growth of carbon nanoparticles, and good agreement on the size of particles with results of in-situ measurements[82] was obtained. Initial size of nucleated particles is very small (containing tens of

atoms[66]) compared to grown particles and can be neglected. With this simplification, a relation for average diameters of the nanoparticles is[43]:

$$D = (2r_0)^{9/5} n_0^{2/5} \left( \frac{4\pi kT}{m_B} \right)^{1/5} t^{2/5}. \qquad (10)$$

Here, $t$ is time elapsed from the beginning of particles agglomeration, $r_0 \approx 1.2 \times 10^{-10} m$ is Wigner-Seitz radius for boron describing distance between atoms in a boron droplet, $n_0$ is density of boron atoms in the gas at the beginning of condensation, $k$ is Boltzmann constant, $m_B$ is mass of boron atom and $T$ is temperature.

Constant temperature is assumed in (10), which is not exactly the case for a cooling gas, however, typically temperature variation between nucleation of boron droplets and formation of BNNTs is rather small (about 20%-30%), and temperature is raised to a small power (10). Hence, temperature variation can be neglected for simplicity, and average temperature $T = 0.5(T_{nucl} + T_{end})$ can be used in (10) resulting in a very small error. Here, $T_{nucl}$ is predicted temperature at which nucleation of boron droplets takes place and agglomeration initiates, which can be predicted by thermodynamics for a gas mixture with a given molar fraction of boron (300 K should be subtracted to account for the nucleation energy barrier). $T_{end}$ is final temperature at which size of the nanoparticles should be determined. With a constant cooling rate $\dot{T}_0$, (10) can be transformed:

$$D = (2r_0)^{9/5} n_0^{2/5} \left( \frac{2\pi k(T_{nucl} + T_{end})}{m_B} \right)^{1/5} \left( \frac{T_{nucl} - T_{end}}{\dot{T}_0} \right)^{2/5} \qquad (11)$$

$T_{end}$ can be taken equal to temperature at which formation of BNNTs and other solid BN structures takes place, $T_{BNNT}$, if size of boron droplet upon formation of BNNTs is of interest. Alternatively, $T_{end}$ can be taken equal to the temperature of boron solidification, $T_{B\_solid}$, if final size of boron droplets is of interest.

In the ICP reactor[22], pressure is 1atm. and molar fraction of boron is about 1% yielding nucleation temperature of $3500 K$ and solid BN formation temperature $T_{BNNT} = 2800 K$, according to the thermodynamic computations (see Figures 2 and 4). Temperature of boron solidification is $T_{B\_solid} = 2300 K$ [39]. Due to energy barrier for the nucleation, actual nucleation temperature is $300 K$ lower than one predicted by the thermodynamics: $T_{nucl} = 3200 K$. For these temperature and boron molar fraction, density of the boron vapor is: $n_0 \approx 3 \times 10^{22} m^{-3}$. Cooling rate $\dot{T}_0$ is about $6 \times 10^4 K/s$ (according to Ref. [22], flow velocity is about 20 m/s and temperature gradient is 3000 K/m in the region of interest) in the case with hydrogen addition, when BNNTs are produced. Substitution of these parameters into (11) yields diameter of the boron droplets upon formation of BNNTs of about $20nm$. It is difficult to validate this value experimentally, *in-situ* measurements diameter of boron droplets diameters amidst the flight, when nanotubes grow on them, would be required. However, validation of this model can be performed by comparing the final size of boron droplets predicted by (11) to *ex-situ* TEM images of boron droplets with no nanotubes produced in the case without addition of hydrogen (see Figure 3f in Ref. [22]). In the no-hydrogen case, cooling rate was about $10^5\ K/s$ (velocity 20 m/s and temperature gradient about 5000 K/m). Substitution of this value and $T_{B\_solid}$ in (11) yields final diameter of boron droplets of about $20nm$ (coincidently, the same as for boron droplets with nanotubes grown with the addition of hydrogen). As seen from the TEM image[22], there were mainly boron particles of $15nm-20nm$, which is in a good agreement with our theoretical prediction.

The same method can be applied to assess size of boron droplets in the electric arc reactor for BNNT synthesis[15]. Unfortunately, no quantitative data is available on boron molar fraction and gas cooling rate within the nanoparticle growth region in electric arc in nitrogen for synthesis of BNNTs. However, for general assessment, analogy with a carbon arc studied in Ref. [43] can be employed here: both arcs have similar size and power, in both arcs one of the electrodes ablates, and evaporated material (boron or carbon) diffuses in surrounding gas to the nanoparticles growth region. This approach makes sense because the results of (11) are weakly dependent on the input parameters. In Ref. [43], simulations predicted coincidently the same parameters as in ICP system[22]: molar fraction of the ablated material in the growth region is about 1%, and cooling rate $\dot{T}_0$ about $2.5\times 10^5\ K/s$. This yields diameter of boron droplets of about $10nm-15nm$. This result is in agreement with *ex-situ* TEM image of boron droplets with nanotubes grown on them (see Figure 3 in Ref. [15]). Again, these estimates give just an order of magnitude, for more accurate predictions, modeling of the arc in nitrogen for BNNT synthesis is required.

### III.G. Analytical relations for species densities

In the thermodynamic approach, densities of species are determined via minimization of relation (7) with constraints (5) and (6). Note that in (7), densities of various species are substantially different, by many orders of magnitude. Apparently, species with higher densities have higher impact on chemical composition of other species within the system. Having this in mind, relation (7) can be substantially simplified when determining density of any particular species. For any species of interest, only species with major densities having the same chemical elements need to be considered.

For instance, when considering atomic nitrogen, only diatomic nitrogen $N_2$ needs to be kept in equation (7) as a dominant nitrogen-containing species. This yields following relation for densities of N and $N_2$:

$$\sum_{i \in N, N_2} N_i \left( kT \left( \ln N_i - \ln \left( \sum_{k \in N, N_2} N_k \right) + \ln \frac{p}{p_0} \right) + G_i(p_0, T) \right) \to \min, \quad (12)$$

with following constraint:

$$N_N + 2N_{N_2} = N_N^*. \quad (13)$$

Here, $N_N$ is number if nitrogen atoms, $N_{N_2}$ is number of nitrogen diatomic molecules, and $N_N^*$ is total number of nitrogen atoms in all species within the system (is constant), $p_0 = 1\,atm.$ is a reference pressure for which the Gibbs energies of formation $G_i^f$ are determined for various species $i$, as introduced in eq. (2).

Solution of (12) and (13) yields following relation:

$$\frac{N_{N_2}(N_{N_2} + N_N)}{N_N^2} = \frac{p}{p_0} \exp\left(\frac{\Delta G_N}{RT}\right),$$

where $\Delta G_N = 2G_N^f(p_0, T) - G_{N_2}^f(p_0, T)$; $G_N$ and $G_{N_2}$ are energies of formation for N and $N_2$ respectively. Or, in terms of species densities (using relation (8)):

$$\frac{n_{N_2}(n_{N_2} + n_N)}{n_N^2} = \frac{p}{p_0} \exp\left(\frac{\Delta G_N}{RT}\right).$$

This relation is known as the law of mass action[46,83] for a dissociation reaction. Taking into account that $N_2$ is the major component in the mixture within temperature range considered, and its partial pressure is roughly equal to the total gas pressure $p$, the following relation for partial pressure of atomic nitrogen is obtained:

$$p_N = \sqrt{p_0 p} \exp\left(-\frac{\Delta G_N}{2RT}\right). \quad (14)$$

As for boron, major boron containing species within the mixture is either its atomic gas phase at higher temperatures or its liquid phase at lower temperatures. In order to find densities of these two species, equation (7) can be simplified to:

$$N_B \left( RT \ln \frac{p}{p_0} + G_B \right) + N_{B,l} G_{B,l} \to \min, \quad (15)$$

with following constraint:

$$N_B + N_{B,l} = N_B^*. \quad (16)$$

Here, subscripts $B$ and $B,l$ refer to atoms in gas and liquid states respectively. Solution of (15) and (16) yields following relation for partial pressure of atomic boron gas:

$$p_B = p_0 \min\left( \exp\left(-\frac{\Delta G_B}{RT}\right), x_{B,\max} \right). \tag{17}$$

Here, $\Delta G_B = G_B^f(p_0, T) - G_{B,l}^f(T) + RT$, and $x_{B,\max} \approx 2b_B$ is maximum molar fraction of boron gas (when all boron is in gas phase). Factor "2" before $b_B$ is to take into account that boron is mostly present in atomic form and nitrogen – in form of $N_2$ molecules:

$$x_{B,\max} \approx \frac{N_B}{N_B + N_{N_2}} \approx \frac{N_B^*}{N_B^* + N_N^*/2} = \frac{2b_B}{1+b_B} \approx 2b_B.$$

Relation (17) is similar to a known Clausius-Clapeyron equation for saturation pressure, but formulated in terms of Gibbs free energies.

Knowing densities and pressures of major nitrogen ($N_2$) and boron (B) gas species allows determining densities of B-N compound molecules $B_x N_y$. For such molecules, equation (7) can be written in a simplified form similar to (12), with summation over three species: $B$, $N_2$ and $B_x N_y$. Following constraints on the species densities should be applied:

$$\begin{aligned} N_B + x N_{B_x N_y} &= N_{B,g}^*, \\ 2 N_{N_2} + y N_{B_x N_y} &= N_N^*. \end{aligned} \tag{18}$$

Here, $N_{B,g}^*$ is total number of boron atoms in the gas phase, which is determined by condensation of boron and, hence, is conserved on conversion of $N_{B_x N_y}$ into other gas species. Both $N_N^*$ and $N_{B,g}^*$ can be considered constants in (18). Solution of this system in a general case is rather bulky; however, taking into account that $N_{B_x N_y} \ll N_N^*$ and usually $N_{B,g}^* < N_N^*$, compact solution can be written:

$$N_{B_x N_y} = \left(N_B^*\right)^x \left(\frac{N_N^*}{2}\right)^{1-x} \left(\frac{p}{p_0}\right)^{\left(\frac{y}{2}+x-1\right)} \exp\left(-\frac{\Delta G}{RT}\right),$$

or, formulated in terms of partial pressures, taking into account that $N_2$ is major component in the mixture:

$$p_{B_x N_y} = \left(p_B\right)^x \frac{p^{\frac{y}{2}}}{p_0^{\frac{y}{2}+x-1}} \exp\left(-\frac{\Delta G}{RT}\right). \tag{19}$$

Here, $\Delta G = G^f_{B_xN_y}(p_0,T) - xG^f_B(p_0,T) - \frac{y}{2}G^f_{N_2}(p_0,T)$. $p_B$ in (19) can be taken as predicted by (17). For temperatures below saturation point of boron this yields:

$$p_{B_xN_y} = \frac{p^{\frac{y}{2}}}{p_0^{\frac{y}{2}-1}} \exp\left(-\frac{\Delta \tilde{G}}{RT}\right), \quad (20)$$

where $\Delta \tilde{G} = G^f_{B_xN_y}(p_0,T) - \frac{y}{2}G^f_{N_2}(p_0,T) - xG^f_{B,l}(T) - xRT$.

Eq. (20) can be considered as a generic equation for boron and nitrogen containing species N, BN, B$_2$N, B, B$_2$ and N$_2$ corresponding to different values of $x$ and $y$. Typically, $\Delta \tilde{G}$ has linear dependence on temperature. In this regard, Eq. (20) can be rewritten in Arrhenius form:

$$p_{B_xN_y} = p^{\frac{y}{2}} p_0^{1-\frac{y}{2}} A_{B_xN_y} \exp\left(-\frac{T_{B_xN_y}}{T}\right). \quad (21)$$

Similar relations can be derived for hydrogen-containing species H, B$_x$H$_y$ and N$_x$H$_y$, for temperatures below 3200 K, where H$_2$ is major hydrogen-containing component (see Figures 3, 4). For convenience, let's formulate it in a general form, for species B$_x$N$_y$H$_z$ (x, y and z can be zero). The relation for the partial pressure as a function of Gibbs free energy change takes a form:

$$p_{B_xN_yH_z} = p^{\frac{y+z}{2}} p_0^{1-\frac{y+z}{2}} \exp\left(-\frac{\Delta \hat{G}}{RT}\right),$$

where $\Delta \hat{G} = G^f_{B_xN_yH_z}(p_0,T) - xG^f_{B,l}(T) - \frac{y}{2}G^f_{N_2}(p_0,T) - \frac{z}{2}G^f_{H_2}(p_0,T) - xRT$.

This yields following Arrhenius form for the partial pressure of B$_x$N$_y$H$_z$ species:

$$p_{B_xN_yH_z} = p^{\frac{y+z}{2}} p_0^{1-\frac{y+z}{2}} A_{B_xN_yH_z} \exp\left(-\frac{T_{B_xN_yH_z}}{T}\right), \quad (22)$$

Values of coefficients $A_{B_xN_yH_z}$ and $T_{B_xN_yH_z}$ for various species are summarized in Table 2.

| Species, $B_xN_yH_z$ | $A_{B_xN_yH_z}$ | $T_{B_xN_yH_z}$, K |
|---|---|---|
| B | 515 000 | 58 800 |
| $B_2$ | 1 962 000 | 81 900 |
| $B_3$ | 238 000 | 74 100 |
| N | 3 600 | 58 100 |
| $N_3$ | $2.9 \times 10^{-3}$ | 51 700 |
| BN | 359 000 | 65 000 |
| $B_2N$ | 535 000 | 40 700 |
| H | 1 570 | 27 500 |
| BH | 5 300 | 44 000 |
| $BH_2$ | 5.2 | 15 400 |
| $BH_3$ | $7.2 \times 10^{-4}$ | 2 780 |
| NH | 11.5 | 45 400 |
| $NH_2$ | $7.1 \times 10^{-3}$ | 22 400 |
| $NH_3$ | $9.7 \times 10^{-7}$ | -6 100 |

*Table 2. Coefficients $A_{B_xN_yH_z}$ and $T_{B_xN_yH_z}$ for various species in Eq. (22) for the species densities.*

Pressures of B atoms, $N_2$, BN and $B_2N$ molecules obtained with formulas (14), (17) and (19) are plotted in Figure 7 as functions of temperature, in comparison with results of full thermodynamic solution for the B-N mixture. As seen from the Figure, at temperatures above the point of solid BN formation, simple algebraic relations (14), (17) and (19) provide pretty accurate values without the need to perform numerical minimization of relation (7). It enables quick analysis of the mixture composition and determination which species are present when BNNTs and other BN structures start to form.

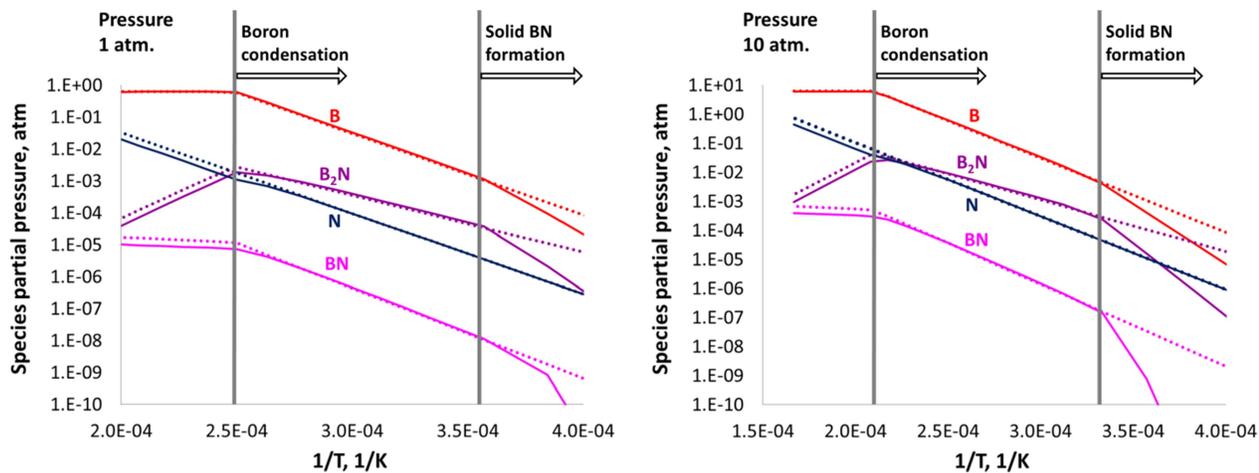

*Figure 7. Partial pressures of selected species obtained using algebraic relations (14), (17) and (19) as functions of temperature (dotted lines), in comparison with results of full thermodynamic modeling of the B-N mixture (solid lines). 45% boron and 55% nitrogen within the mixture, two pressures are considered. Good agreement between full modeling results and analytical solutions is observed for temperatures higher than point of solid BN formation.*

## Conclusions

Calculations of equilibrium chemical composition for B-N mixture with and without addition of hydrogen at 1 atm. and 10 atm. in the temperature range between 1000K and 6000K were performed using thermodynamic approach. A broad set of species was considered including BN, $B_2N$ and $BN_2$ gas phase molecules, liquid boron and solid BN. Latest allowed accounting for condensation of boron and determining conditions at which formation of solid BN structures take place, which can be interpreted as conditions for growth of BNNTs. The thermodynamic calculations predict that in a cooling mixture, boron condenses at first providing seed particles for the growth of BNNTs, supporting "root growth" mechanism. It was shown that condensation of boron has drastic effect on the gas phase composition yielding reduction of boron-containing species densities by several orders of magnitude at temperatures corresponding to the formation of boron liquid (droplets) and BNNTs. This effect was not considered in previous thermodynamic studies of B-N mixtures and should be taken into account when studying conditions for growth of BNNTs.

Results of the thermodynamic calculations have shown that $B_2N$ molecules have the highest density among nitrogen-containing species, after $N_2$, when BNNTs grow on surface of boron droplets. Unlike $B_2N$ molecule, $N_2$ has high dissociation energy and can hardly dissociate on a boron surface in order to contribute to the formation of periodic B-N structure within BNNT. Hence, $B_2N$ molecules can be suggested as major source of nitrogen and major building molecule for formation and growth of BNNTs. These predictions are supported with experimental results obtained by our OES measurements during laser ablation of boron-rich target. Atomic nitrogen can give some contribution to the growth as well. Other nitrogen-containing species have much lower densities and cannot have substantial impact on the growth of BNNTs.

The results of the thermodynamic calculations were also verified by qualitative comparison to spectroscopic measurements of the gas composition at several locations within the ICP plasma reactor of Ref. [22]. Measured spectral line intensities of several species (N, $N_2$, NH, B, BH, H, BN, $H_2$) at three locations in the ICP plasma reactor qualitatively agree with predictions of the thermodynamic calculations.

Effect of the pressure increase from 1 atm. to 10 atm. on the gas composition was studied. It was shown that because density of boron gas is determined merely by the saturation curve and does not depend on the background pressure, densities of N, BN and $B_2N$ molecules increase proportionally to square root of pressure. These results suggest that precursors for formation of BNNTs remain the same ($B_2N$ molecules and N atoms), and enhanced production yield and purity of BNNTs observed in Refs. [17] and [20] at higher pressures can be attributed to increase of densities of these precursors.

Convenient analytical relations in Arrhenius form (22) were derived to describe species densities at various pressures and temperatures higher than point of solid BN formation. Values of coefficients for the Arrhenius expression for various species were provided.

References [21] and [22] observed enhanced BNNT production with hydrogen addition. Our analysis shows that formation of $NH_3$ molecules at temperatures of BNNT growth might be responsible for this effect. Nitrogen atoms react fast with abundant $H_2$ molecules to form $NH_2$ and then $NH_3$ species, instead of just recombining back to inert $N_2$ molecules in the no-hydrogen case. These $NH_3$ molecules can provide additional nitrogen to boron droplets and hence enhance BNNT growth.

We calculate the size of boron droplets where BNNTs grow. Boron gas density, temperature and flow velocity were taken from the fluid modeling[22,43]. Good agreement between our simulation results and observed in *ex-situ* TEM photos[15,22] for the droplets size was obtained. It was shown that the initial fraction of boron in the system does not affect composition of the gas mixture when BNNTs grow, because most of boron condenses into liquid droplets (though the initial fraction of boron affects size of boron droplets on which the BNNTs grow, see Eq. (10)).

# Appendix A. Details on the experimental setup for laser ablation of boron-rich targets and OES measurements

Laser ablation is driven by Nd:Yag laser (Surelite III-10), equipped with second and third harmonic generators. The laser produced pulses of 1064 nm wavelength at frequency of 10 Hz, which are converted to 355 nm, by the frequency conversion unit. The emerging laser has a full-width-at-half-maximum (FWHM) duration of ~7 ns. The laser beam was guided into the investigation area and focused on the target with lens (f=300 mm). The target is boron-nitride (BN) tablet, having diameter of 25 mm and a height of 5 mm. The BN target was held in a 3D-printed pedestal, which in turn was connected to a rotation motor. Fast rotation of the target prevents the dipping of the laser into the target. The resulting ablation plume is imaged with a lens on the entrance of the imaging spectrometer (iHoriba 550). The entrance slit of the spectrometer was set to a width of 50 microns and the spectrometer grating of 1200 groves/mm was employed to diffuse the light. An iCCD camera (PI-MAX 3, 1024x1024 pixel CCD) is connected to the exit port of the spectrometer for detection. A Dove prism located between the lens and the spectrometer slit rotates the image by $90^0$. In this fashion the ablation plume was imaged on the iCCD vertically: the target was on the top of the CCD, the plume is extended from top to bottom and horizontal direction on the iCCD corresponds to different wavelengths.

# Appendix B. Assessment of the plasma decay time at the surfaces of boron droplets

In ionized gas, particles coagulation might be affected by their charging, as shown e.g. in Refs. [79], [80]. However, in these references, low pressure high temperature (5 eV) plasmas were considered, with ionization rate sufficient to compensate the loss of charge particles due to the ion (positive ion-electron) recombination at the droplets surfaces and maintain the plasma. In atmospheric (and higher) pressure nanosynthesis reactors, at temperature of boron condensation and coagulation (about 0.3 eV), ionization rate is much lower, and plasma decay on surfaces of boron droplets plays important role.

At these temperatures, effects of photo- and thermal emission on the particles charging are negligible. The droplets are charged negatively due to bombardment by plasma electrons. These negatively charged droplets attract ions that recombine on their surfaces neutralizing the droplet charge (boron is conductive in a liquid state[84]). Volumetric ion recombination rate at the droplets surfaces can be assessed as:

$$r_r \geq \pi d^2 n_d n_i \sqrt{\frac{2kT}{m_B}} \,. \qquad (B.1)$$

Here, $d$ is droplets' dimeter, $n_d$ and $n_i$ are densities of boron droplets and ions respectively, $m_B$ is mass of boron ion. This is an estimate from below; effect of the droplets charge is neglected for simplicity (negative charge of the droplets only increases the recombination rate).

For convenience, density and diameter of boron droplets can be linked via mass conservation:

$$n_d = 8 n_0 r_0^3 / d^3 . \tag{B.2}$$

Here $n_0 \approx 3 \times 10^{22} \, m^{-3}$ is density of boron atoms before the condensation, $r_0 \approx 1.2 \times 10^{-10} \, m$ is Wigner-Seitz radius (see Section III.F).

Substitution of this relation into Eq. (B.1) yields for the plasma decay time:

$$t_{decay} \leq \frac{d}{8 \pi n_0 r_0^3} \sqrt{\frac{m_B}{2kT}} . \tag{B.3}$$

According to this expression, plasma decays quite fast in the beginning of coagulation process, when the droplets are small. For 1 nm droplets, plasma decay time is $t_{decay} \leq 5 \cdot 10^{-7} s$. This is considerably faster than the coagulation time ($10^{-5} s$ for droplets to reach 1 nm in diameter, according to Eq. (10) from Section III.F).

Volumetric ionization rate is given by:

$$r_i = k_i n_e n_B . \tag{B.4}$$

Here, $k_i$ is a rate coefficient, $n_e$ and $n_B$ are densities of electrons and boron atoms. For boron at temperature 0.3 eV, rate coefficient is $k_i = 10^{-19} \, m^3 / s$, calculated using formula (59) from Ref. [85] for step-ionization of an atom at low temperatures. Taking into account that $n_e < n_i$ (negative charge from plasma is partially accumulated on the droplets) and $n_B < n_0$, estimate from below for the ionization time is:

$$t_{ioniz} \geq (k_i n_0)^{-1} = 3 \cdot 10^{-4} s . \tag{B.5}$$

This is much larger than the recombination time ($t_{decay} \leq 5 \cdot 10^{-7} s$, as discussed earlier); hence, effect of ionization in the plasma is negligible.

From these assessments it is clear that as soon as the droplets appear, the plasma will virtually immediately disappear (much faster than solid BN formation initiates after about 0.01 s). Negative charge will be removed from the droplets via recombination of ions, and the coagulation will take place for uncharged droplets, as described in Section III.F.

# Appendix C. Applicability of the thermodynamic approach to model the nanosynthesis systems.

Thermodynamics is a useful theoretical tool for determining equilibrium composition of complex reacting systems of multiple components without the need to go in details of a chemical mechanism that lead a system to this state. The latest would require quite extensive calculations of chemical reactions rates among various species within a system; rate coefficients for some of these reactions are often unknown.

If physical timescale for variation of external conditions (temperature, pressure etc.) is greater than a chemical timescale at which the mixture composition adapts to the conditions change, then the system can be considered equilibrium at each timeframe, and the thermodynamic approach is valid. If rates of some species formation/destruction are low, then densities of these species can deviate from an equilibrium solution.

The aim of this paper is to find out what gas phase species bring nitrogen to surfaces of boron droplets when BNNTs grow on them. Molecular nitrogen ($N_2$), most abundant species in the system, cannot be directly built into a periodic B-N structure of BNNTs. $N_2$ molecules have a strong bond (9.8 eV) and do not dissociate on boron droplets surfaces[26]. However, $N_2$ molecules can be efficiently dissociated in a gas volume via stepwise excitation of its vibrational degree of freedom[86,87] through collisions with gas phase species, ultimately resulting in the bond breakup. Atomic nitrogen released in this process can participate in growth of BNNTs by either directly impinging boron droplets or via recombination with other gas phase species to form molecules with relatively low binding energies that can release atomic nitrogen (or a B-N couple) at a boron droplet surface.

The thermodynamic modeling predicts that $B_2N$ and then (at lower temperature) $NH_3$ molecules have the highest density among nitrogen-containing species in a temperature range corresponding to BNNTs growth. BN and NH molecules, which has been suggested in Ref. [22] as potential channels to provide nitrogen for BNNTs, have, according to the thermodynamic results, noticeable densities only at higher temperatures (above 4000 K) which then rapidly decay when temperatures decrease to the values corresponding BNNTs growth (≤2800 K).

In order to verify whether thermodynamic method can be applied to describe nano-synthesis systems, timescales for $B_2N$ and $NH_3$ formation and for BN and NH destruction (dissociation) have to be compared to a timescale of temperature variation in the system. The latest is about 0.01 s in the ICP/arc discharge reactors (temperature decay rate is $10^5$ K/s, temperature variation scale is 1000 K).

For the sake of simplicity, when assessing timescale of $B_2N$ and $NH_3$ formation, only one chemical pathway for each of these molecules will be considered. Additional formation mechanisms can only increase the formation (production) rates (reduce formation time).

Pathway for $B_2N$ molecules formation is:

$N_2 + M \rightarrow N + N + M$;

$N + B + M \rightarrow BN + M$;

$B + BN + M \rightarrow B_2N + M$.

Pathway for $NH_3$ molecules formation is:

$N + H_2 + M \rightarrow NH_2 + M$;

$NH_2 + H_2 \rightarrow NH_3 + H$.

$N_2$ and $H_2$ molecules and B atoms are major nitrogen- and boron-containing components within the system, present in abundance. "M" represents all components in the system.

Production rate of a final product ($B_2N$ / $NH_3$) is a minimum of production rates at each step of a chemical mechanism. Assessments of the production rates are summarized in Tables C.1 and C.2. Reaction rates for BN and $B_2N$ molecules recombination were assessed using a simple theory, formula (6.27) from Ref. [87].

| Chemical reaction | Temp., K | Rate coef. | Ref. | Reactants densities, $m^{-3}$ (according to thermod.) | Production rate, $m^{-3}$/s |
|---|---|---|---|---|---|
| $N_2 + M \rightarrow N + N + M$ | 4000 | $10^{-26}$ $m^3$/s | 88 | $N_2$: $10^{24}$<br>B: $10^{24}$ | $4 \cdot 10^{22}$ |
| $N + B + M \rightarrow BN + M$ | | $10^{-44}$ $m^6$/s | 87 | N: $2 \cdot 10^{21}$<br>BN: $10^{19}$<br>M: $2 \cdot 10^{24}$ | $4 \cdot 10^{25}$ |
| $B + BN + M \rightarrow B_2N + M$ | | | | | $2 \cdot 10^{23}$ |

Table C.1. Production rates within a pathway for $B_2N$ molecules formation.

Density of $B_2N$ molecules is at its highest level at 4000 K (see Fig. 3); production rates in Table C.1 were calculated for this temperature. As it can be seen from the Table, the slowest reaction in the $B_2N$ formation mechanism is the dissociation of nitrogen (recombination reactions to form BN and then $B_2N$ go faster). This reaction might be a "bottleneck" limiting formation of $B_2N$ molecules. The production rate of $4 \cdot 10^{22}$ $m^{-3}$/s determined by the $N_2$ dissociation (two N atom are produced per each dissociation event) is sufficient to raise density of $B_2N$ molecules to $4 \cdot 10^{20}$ $m^{-3}$ (molar fraction $2 \cdot 10^{-4}$) on a timeframe of 0.01 s (during which temperature in the system is close to the considered value). This density value is about five times lower than the peak density predicted by the thermodynamics, but it is considerably higher than the density at temperature when BNNTs grow. Thereby, for a mixture cooling in a reactor, the "peak" of the $B_2N$ density profile can be slightly "cut off" on the top, but density at the BNNTs growth should agree with the thermodynamics predictions.

Note that, in the nanosynthesis reactors, the mixture comes from a hot zone (temperatures above 6000 K) where nitrogen is highly dissociated. Hence, there might be already sufficient amounts of atomic nitrogen in the system. With this, the system might "bypass" the nitrogen dissociation bottleneck and go straight to formation of BN and $B_2N$ molecules via fast recombination reactions. In this regard, even peak $B_2N$ density value might be in a good agreement with the thermodynamic prediction. This question is not important from the standpoint of BNNTs growth but might be interesting from a theoretical point of view. Detailed chemical kinetic calculations accounting for simultaneous act of multiple reactions would be required to answer this question. This is a complex task additionally complicated by unavailability of rate coefficients for many reactions. Such considerations are out of the scope of this paper.

| Chemical reaction | Temp., K | Rate coef. | Ref. | Reactants densities, $m^{-3}$ (according to thermod.) | Production rate, $m^{-3}/s$ |
|---|---|---|---|---|---|
| $N_2+M \rightarrow N+N+M$ | 2800 | $2 \cdot 10^{-31}$ $m^3/s$ | 88 | $N_2$: $10^{24}$ $H_2$: $10^{24}$ | $8 \cdot 10^{17}$ |
| $N+H_2+M \rightarrow NH_2+M$ | | $10^{-44}$ $m^6/s$ | 89 | $NH_2$: $10^{18}$ $M$: $2 \cdot 10^{24}$ | $2 \cdot 10^{24}$ |
| $NH_2+H_2 \rightarrow NH_3+H$ | | $10^{-17}$ $m^3/s$ | 90 | | $10^{25}$ |

*Table C.2. Production rates within a pathway for $NH_3$ molecules formation.*

According to the thermodynamic prediction (Fig. 3), $NH_3$ molecules become major nitrogen-containing gas-phase component at temperature of BNNTs formation (2800 K), with exponentially raising density (with the temperature decay). As seen from Table C.2, nitrogen dissociation reaction is the slowest one in the $NH_3$ production pathway at this temperature. However, since a gas mixture comes from a hotter reactor region, it already has decent amount of atomic nitrogen. According to the thermodynamics results (Fig. 3), amount of atomic nitrogen in the mixture at temperature 3000 K (right before $NH_3$ molecules are produced) exceeds the amount of $NH_3$ molecules produced at 2800 K. Within the mixture, nitrogen atoms recombine with each other (to form $N_2$ molecules) and with $H_2$ (to form $NH_2$ molecules with their further transformation to $NH_3$). Density of $H_2$ molecules is by many orders of magnitude higher than density of N atoms at this temperature. Rate of $NH_2+H_2$ chemical reaction leading to $NH_3$ molecules formation is very high (Table C.2). Thereby, almost all atomic nitrogen should be incorporated, very quickly, in $NH_3$ molecules, which should reach densities predicted by the thermodynamics on a timescale of $10^{-5}$ s. It means, much faster than temperature variation timescale. In other words, presence of hydrogen in the system should prevent recombination of N atoms into $N_2$ molecules due to their much faster recombination with abundant molecular $H_2$ to form $NH_3$ molecules. These $NH_3$ might serve as efficient source of nitrogen enhancing growth of BNNTs.

Rates and timescales of BN and NH dissociation are presented in Table C.3. Dissociation rate coefficient for BN molecule was calculated using an equilibrium constant and formula (6.27) from Ref. [87] for a reversed reaction. As seen from the table, dissociation of these molecules is much faster (timescale $5 \cdot 10^{-4}$ s) than temperature variation (timescale 0.01 s). Thereby, thermodynamic prediction should accurately predict densities of these molecules at temperatures of BNNTs growth, showing superior densities of $B_2N$ and $NH_3$ molecules.

| Chemical reaction | Temp., K | Rate coefficient | Ref. | Reactants densities, $m^{-3}$ | Characteristic time, s |
|---|---|---|---|---|---|
| NH+M→N+H+M | 3000 | $10^{-21}$ $m^3$/s | 91 | M: $2 \cdot 10^{24}$ $m^{-3}$ | $5 \cdot 10^{-4}$ s |
| BN+M→B+N+M | 3000 | $10^{-21}$ $m^3$/s | 87 | M: $2 \cdot 10^{24}$ $m^{-3}$ | $5 \cdot 10^{-4}$ s |

*Table C.3. Rates and timescales of molecules dissociation.*

## Acknowledgements

The authors would like to thank Dr. Predrag Krstic (Stony Brook University), Dr. Longtao Han (Stony Brook University), Dr. Rachel Selinsky (Princeton University), Dr. Roberto Car (Princeton University), Dr. Biswajit Santra (Princeton University) and Dr. Andrei Khodak (PPPL) for fruitful discussions. The thermodynamic modeling was supported by the US Department of Energy (DOE), Office of Science, Fusion Energy Sciences. Laser ablation experiments and nanoparticle growth calculations were supported by the US DOE, Office of Science, Basic Energy Sciences, Materials Sciences and Engineering Division.